\documentclass[twocolumn,prl,showpacs,preprintnumbers,amssymb]{revtex4-1}

\usepackage{epsfig}
\usepackage{dcolumn}
\usepackage{bm}

\usepackage{amsmath}
\usepackage{amssymb}
\usepackage{latexsym}
\usepackage{epsfig}
\usepackage{color}
\usepackage{mathtools}
\usepackage{youngtab}
\usepackage{bigints}
\usepackage{sidecap}

\def\z2z2{$\IC^3/(\IZ_2\times\IZ_2)$}

\def\id{{\bf 1}}

\def\cF{{\cal F}}

\def\cK{{\cal K}}

\def\cM{{\cal M}}

\def\cV{{\cal V}}

\def\a{\alpha}

\def\b{\beta}

\def\d{\delta}\def\D{\Delta}

\def\k{\kappa}
\def\l{\lambda}

\def\p{\pi}

\def\s{\sigma}

\def\th{\theta}

\def\beq{\begin{equation}}\def\eeq{\end{equation}}
\def\beqa{\begin{eqnarray}}\def\eeqa{\end{eqnarray}}
\def\barr{\begin{array}}\def\earr{\end{array}}

\def\wt{\widetilde}

\def\ds {{\del \hspace{-6.4pt} \slash}\;}

 \let\br=\bigr

\def\bd{\begin{document}}
\def\ed{\end{document}}
\def\ba{\begin{array}}
\def\ea{\end{array}}
\def\bea{\begin{eqnarray}}
\def\eea{\end{eqnarray}}
\def\ft#1#2{{\textstyle{{\scriptstyle #1}\over {\scriptstyle #2}}}}
\def\fft#1#2{{#1 \over #2}}
\newcommand{\be}{\begin{equation}}
\newcommand{\ee}{\end{equation}}
\newcommand{\eq}[1]{(\ref{#1})}
\def\eqs#1#2{(\ref{#1}-\ref{#2})}
\def\det{{\rm det\,}}
\def\tr{{\rm tr}}
\newcommand{\ho}[1]{$\, ^{#1}$}
\newcommand{\hoch}[1]{$\, ^{#1}$}
\def\ra{\rightarrow}

\def\Xh{\hat{X}}
\def\ah{\hat{a}}
\def\xh{\hat{x}}
\def\yh{\hat{y}}
\def\ph{\hat{p}}
\def\G{{\cal G}}
\def\Dth{{\Delta_\th}}

\def\bk{{\bf k}}
\def\bx{{\bf x}}
\def\br{{\bf r}}
\def\tr{{\rm tr \,}}
\def\Tr{{\rm Tr \,}}
\def\diag{{\rm diag \,}}
\def\tg{{\rm tg \,}}
\def\ov{\overline}

\def\preal{{\rm Re\,}}
\def\pim{{\rm Im\,}}

\def\ds{\displaystyle}
\def\yzero{\smash{\hbox{$y\kern-4pt\raise1pt\hbox{${}^\circ$}$}}}
\def\p{\partial}
\def\a{\alpha}
\def\b{\beta}
\def\g{\gamma}
\def\d{\delta}
\def\beq{\begin{equation}}
\def\eeq{\end{equation}}
\def\beqa{\begin{eqnarray}}
\def\eeqa{\end{eqnarray}}
\def\Om{\Omega}
\def\om{\omega}
\def\th{\theta}
\def\vt{\vartheta}
\def\vphi{\varphi}
\def\-{\hphantom{-}}
\def\ov{\overline}
\def\s2{\frac{1}{\sqrt2}}
\def\wh{\widehat}
\def\wt{\widetilde}
\def\oh{\frac{1}{2}}
\def\tr{{\rm tr \,}}
\def\Tr{{\rm Tr \,}}
\def\diag{{\rm diag \,}}
\def\vac{|0 \rangle}
\def\vm{\relax{n_{\text{v}}}}
\def\tv{\tilde v}
\def\Dsl{\,\raise.15ex\hbox{/}\mkern-13.5mu D} 
\def\id{{\rm 1}}

\def\ti{\times}
\def\til{\tilde}
\def\eps{\epsilon}
\def\k{\kappa}
\def\A{\Arrowvert}
\def\cw{{\cal W}}
\def\G{\Gamma}
\def\car{{\cal R}}
\def\l{\lambda}
\def\raw{\rightarrow}
\def\Raw{\Rightarrow}
\def\inte{{\bf Z}}
\def\cpx{{\bf C}}
\def\real{{\bf R}}
\def\Lam{\Lambda}
\def\D{\Delta}
\def\cb{{\cal B}}
\def\ca{{\cal A}}

\def\re{\mbox{Re }}
\def\im{\mbox{Im }}
\def\tr{\mbox{Tr}}
\def\str{\mbox{STr}}

\def\IC{\mathbb{C}}
\def\IN{\mathbb{N}}
\def\IZ{\mathbb{Z}}
\def\IR{\mathbb{R}}
\def\IP{\mathbb{P}}
\def\Id{{\mathbb{I}}}

\begin{document}

\preprint{IFT-UAM/CSIC-15-056}
\preprint{MPP-2015-109}

\title{Large field inflation from D-branes}
\author{Dagoberto Escobar,$^{1}$ Aitor Landete,$^{1}$ Fernando Marchesano,$^{1}$ and Diego Regalado$^{2}$}
\affiliation{\small\slshape  $^{1}$  Instituto de F\'{\i}sica Te\'orica UAM-CSIC, Cantoblanco, 28049 Madrid, Spain \\
$^{2}$Max-Planck-Institut f\"ur Physik, F\"ohringer Ring 6, 80805 Munich, Germany }
\begin{abstract}
We propose new large field inflation scenarios built on the framework of F-term axion monodromy. Our setup is based on string compactifications where D-branes create potentials for closed string axions via F-terms. Because the source of the axion potential is different from the standard sources of moduli stabilisation, it is possible to lower the inflaton mass as compared to other massive scalars. 
We discuss a particular class of models based on type IIA flux compactifications with D6-branes. In the small field regime they describe supergravity models of quadratic chaotic inflation with a stabiliser field. In the large field regime the inflaton potential displays a flattening effect due to Planck suppressed corrections, allowing to easily fit the cosmological parameters of the model within current experimental bounds. 
\end{abstract}
\maketitle

\section{Introduction}

Since put forward in \cite{Linde:1983gd}, large field chaotic inflation has been an attractive proposal for describing early universe cosmology. This has motivated its embedding in more sophisticated schemes, more particularly within models of $N=1$ supergravity and 
superstring theory. However, such embeddings suffer from a number of important subtleties which need to be addressed before claiming to have constructed a successful model of large field inflation. 

A well-known caveat in the context of supergravity models is that the simplest superpotential leading to chaotic inflation, $W_{\rm inf} =  X^2$, results in a scalar potential unbounded from below at large values of the inflaton. This problem is typically addressed by considering the alternative superpotential \cite{Kawasaki:2000yn} (see \cite{Kallosh:2010xz} for generalisations)
\be
W_{\rm inf}\, =\, S X
\label{KYY}
\ee
in which a second chiral multiplet $S$ has been introduced. The r\^{o}le of $S$, dubbed {\em stabiliser} field, is to generate a potential for the axionic scalar within $X$ while keeping a vanishing vacuum expectation value during inflation, therefore controlling the negative term $-3e^K|W|^2$ in the scalar potential for large values of $\langle X \rangle$.

Another central issue is that realistic models of string theory and supergravity contain many other scalar fields beyond those driving inflation. These extra scalar fields must develop a scalar potential on their own that stabilises them above the inflaton mass and the Hubble scale, and in such a way that inflationary dynamics is not disturbed. Finally, the decoupling from the inflaton sector must guarantee that these heavier scalars are not destabilised for trans-Planckian values of the inflaton vacuum expectation value. 

In string theory, a promising scheme to realise large field inflation relies on the idea of axion monodromy, in which an axion develops a potential $V$ that simultaneously breaks its shift symmetry and periodicity \cite{Silverstein:2008sg,McAllister:2008hb}. Particularly interesting for the above discussion are those models classified as F-term axion monodromy inflation \cite{Marchesano:2014mla} (see also \cite{Blumenhagen:2014gta,Hebecker:2014eua}), where for small values of the inflaton field $V$ can be understood as a supergravity F-term potential. Indeed, this supergravity description allows to clearly see the interplay between the stabilised moduli and the inflaton sector, and to determine whether the above or additional subtleties are present in a given model. 

Indeed, the vantage point of supergravity has been recently used to analyse certain F-term axion monodromy inflation models, more precisely type IIB compactifications with fluxes in which the inflaton is a complex structure field \cite{Blumenhagen:2014nba,Hayashi:2014aua,Hebecker:2014kva}. It was found that {\it i)} it is generically very difficult to lower the mass of the inflaton candidate with respect to the other complex structure scalars and that {\it ii)} the inflaton will typically backreact on the closed string moduli. 

While these observations have been used to claim that a successful model of F-term axion monodromy inflation is hard to obtain, it is important to keep in mind that they arise from a very specific setup, and this is where the actual challenge to realise inflation may be rooted in. Indeed, in the setup of \cite{Blumenhagen:2014nba,Hayashi:2014aua,Hebecker:2014kva} the same source (a background flux induced potential) is used to stabilise most complex structure moduli at very high scale and to provide a mass for the inflaton several orders of magnitude below, naturally requiring some fine-tuning. Moreover, the superpotential grows significantly for trans-Planckian values of the inflaton, modifying all the F-terms and creating an important source of moduli destabilisation. 

The purpose of this letter is to point out that these issues can be solved by considering more general setups, in which the sources for moduli stabilisation and inflaton potential are very different. The class of models which we propose do also belong to the framework of F-term axion monodromy inflation, and their main new ingredient compared to previous proposals is a superpotential of the form (\ref{KYY}) involving the inflaton field and generated by the presence of a D-brane filling 4d space-time.

As shown in \cite{Marchesano:2014iea} one achieves a superpotential of the form (\ref{KYY}) by considering space-time filling D-branes with a certain topology in type II orientifold compactifications \footnote{See \cite{Font:2006na,Dudas:2014pva,Hayashi:2014aua} for other proposals to realise the superpotential (\ref{KYY}) in type II compactifications.}. In such setting one of the two complex fields in (\ref{KYY}) is a modulus of the D-brane, while the other is a closed string modulus. 

Now, because this superpotential source is of different nature from a flux-induced superpotential, it can generate masses at a different scale. More precisely, we will see that at the supergravity level the masses generated by the D-brane-induced superpotential depend on the open string kinetic terms, which can be very different from the closed string kinetic terms entering the flux-induced superpotential. In particular, one may generate a hierarchy of masses by considering that the above D-brane is placed in a strongly warped region of the compactification manifold, lowering the inflaton mass with respect to other moduli. 

Of course, if (\ref{KYY}) is the only source of scalar potential for the complex fields $X$ and $S$ one may be faced with a model with four real fields of comparable mass, leading to multifield dynamics during the inflationary period which may in turn generate abundant isocurvature perturbations. There could be however several scenarios which lead to simplifications of this picture, leaving an effective theory with only one or maybe two real fields to drive inflation. For instance:

\begin{itemize}

\item[{\it i)}] It may happen that, out of the four scalars, only one has a shift symmetry in the K\"ahler potential. In particular, one may consider the case where the K\"ahler potential is a function of $S\bar{S}$ and $(\im X)^2$. These assumptions are the ones typically used in the supergravity literature (see e.g., \cite{Kawasaki:2000yn,Kallosh:2010xz}), but we will consider below a string theory construction that satisfies them as well. Then, by the results in \cite{Kallosh:2010xz} the scalar potential will have a minimum at $S = \im X = 0$ and the fields $S$ and $\im X$ will be heavier than $\re X$ and the Hubble scale if the derivatives of the K\"ahler potential satisfy certain inequalities.

\item[{\it ii)}] One of the two complex fields, say $S$, may appear in the piece of the superpotential $W_{\rm mod}$ that implements moduli stabilisation for the scalar fields outside the inflaton sector. It will therefore be fixed together with these moduli at a higher scale than the inflaton mass, leaving an effective theory with only the complex field $X$. Then, if $\re X$ is an axion with the corresponding shift symmetry, $\im X$ will be a saxion which will appear in the K\"ahler potential, and through it in the F-terms of the moduli stabilised at this high scale. It could then be that via these couplings the saxion acquires a high mass during inflation, or perhaps a quartic coupling that stabilises it at a small or vanishing value while inflating with $\re X$. 

\end{itemize}

\section{A type IIA scenario}

Let us render our general discussion more precise by considering a class of string theory models to which the above considerations apply, and that were already suggested in Appendix A of \cite{Marchesano:2014mla}. Let us in particular consider 4d type IIA compactifications with O6-planes and background fluxes \cite{thebook}. As shown in \cite{Marchesano:2014iea}, in this setting one may obtain a superpotential of the form (\ref{KYY}) as follows. We consider a D6-brane wrapping a special Lagrangian three-cycle $\Pi_3$ of the compactification manifold $\cM_6$, such that $b_1(\Pi_3)=1$. This in particular implies that the D6-brane has a complexified position modulus $\Phi$ defined as
\be
A + [\iota_X J_c]_{\Pi_3} \, =\, \Phi\, \zeta 
\ee
with $X$ a normal vector describing a special Lagrangian deformation of $\Pi_3$, $A$ the D6-brane Wilson line profile and $\zeta$ is a harmonic one-form that generates $H^1(\Pi_3, \IZ)$. Finally, $J_c$ is the complexified K\"ahler form $J_c = B + iJ$. Since $b_1(\Pi_3)=1$ the three-cycle $\Pi_3$ will contain a two-cycle $\pi_2$ in the Poincar\'e dual class of $[\zeta]$. We may now assume that $[\pi_2]$ is non-trivial in the homology of $\cM_6$. This implies that
\be
\int_{\pi_2} J_c\, =\, \sum_a n_a T^a\, \equiv\, T
\ee
where $n_a \in \IZ$ describe the topology class of $\pi_2$ and $T^a = b^a + ie^{\phi/2} v^a$ are K\"ahler moduli of the compactification (see \cite{Grimm:2004ua,Grimm:2011dx,Kerstan:2011dy} for conventions). Then, following \cite{Marchesano:2014iea} one can see that the following superpotential is generated
\be
W_{\rm inf}\, =\, \Phi\,  T
\label{Winf}
\ee
where recall that $T$ is a linear combination of K\"ahler moduli. For instance one could have $T = T^1 - T^2$, so demanding $T=0$ (which implies the D6-brane BPS condition $J_c|_{\Pi_3} =0$) does not require any two-cycle of $\cM_6$ shrinking to vanishing size, but rather certain relations among their volumes. 

We would now like to construct a model of inflation from the scalar potential related to (\ref{Winf}). In order to see if this is feasible, one must first understand the interplay between the inflationary potential and the potential fixing the moduli of the compactification. Near the vacuum this can be done by describing the whole system in terms of a 4d supergravity potential
\be
V\, =\, e^K \left(K^{\a\bar{\b}}D_\a W D_{\bar{\b}} \bar{W} - 3|W|^2  \right)
\label{pot}
\ee
where the full superpotential is
\be
W\, =\, W_{\rm mod} + W_{\rm inf}
\label{Wfull}
\ee
with the following moduli stabilisation superpotential 
\be
W_{\rm mod}\, =\, W_{\rm flux}(T,N) + W_{\rm D2}(N,\Phi) + W_{\rm WS}(\Phi,T)\, .
\label{Wmod}
\ee
Here $W_{\rm flux}$ is the superpotential generated by the closed string fluxes threading $\cM_6$, affecting the complexified K\"ahler, complex structure and dilaton moduli. $W_{\rm D2}$ is the superpotential generated by Euclidean D2-brane instantons which not only includes the complex structure moduli of the compactification but also the D6-brane moduli like $\Phi$. Finally, $W_{\rm WS}$ is the correction generated by worldsheet instantons. These instantons can be closed or open Euclidean strings, the former only depending on the K\"ahler moduli of the compactification and the latter also on the D6-brane moduli. 

To evaluate (\ref{pot}) we also need the K\"ahler potential for these fields. From \cite{Grimm:2004ua,Grimm:2011dx,Kerstan:2011dy} we obtain that at tree-level $K = K_K + K_Q$ where
\bea
\label{KK}
K_K &  = & -{\rm log} \left(\frac{i}{6} \cK_{abc} (T^a - \bar{T}^a)(T^b - \bar{T}^b)(T^c - \bar{T}^c) \right) \\ \nonumber
K_Q & = & -2\, {\rm log} \left(\frac{1}{2i} \cF_{{K}{L}} \left[N^{{K}} - \bar{N}^{{K}} + \frac{i}{4} Q^{{K}} \Phi \bar{\Phi} \right]\right. \\ & &\left. \hspace*{2.25cm} \cdot \left[N^{{L}} - \bar{N}^{{L}} + \frac{i}{4} Q^{{L}} \Phi \bar{\Phi} \right] \right)
\label{KQ}
\eea
where $\cK_{abc} \in \IZ$ are triple intersection numbers of $\cM_6$, and we refer the reader to \cite{Kerstan:2011dy} for the definitions involved in (\ref{KQ}). What is more relevant for us is that this K\"ahler potential meets all the symmetry requirements of \cite{Kallosh:2010xz} if we {\it i)} identify $\Phi$ with the stabiliser field $S$ and {\it ii)} choose the triple intersection numbers such that $K_K$ only depends on $(\im T)^2$. In that light, it is natural to identify the inflaton candidate with the axion $\re T$, that is with
\be
b\, =\, \int_{\pi_2} B
\label{binf}
\ee
which is what we will assume in the following.

Since we are aiming to stabilise all moduli fields besides the inflaton at a much higher scale, it make sense to understand moduli fixing as a two-step process. We first forget about the D6-brane and its field $\Phi$,  and only consider the closed string background for which $W = W_{\rm mod}^{\Phi=0}$. We then assume that the we can find a vacuum where the F-terms vanish and almost all moduli are stabilised, with a very small or vanishing value $W^0_{\rm mod}$ for $W_{\rm mod}^{\Phi=0}$ at this vacuum. Because we do not want to stabilise $b$ at this step we will assume that $W_{\rm mod}$ does not depend on $T$, which immediately singles out $b$ as a flat direction of this potential. We then reinstate the D6-brane, hence the field $\Phi$ and the superpotential $W_{\rm inf}$ and see how the full scalar potential computed with (\ref{Wfull}) looks around this sublocus of the moduli space \cite{toappear}.

To see the consequences of this approach let us split the scalar potential (\ref{pot}) as $V = V_Q + V_K - 3e^K|W|^2$, where
\be
V_Q\, =\, e^K \left(K^{\a\bar{\b}}D_\a W D_{\bar{\b}} \bar{W} \right) \quad \quad \a, \b = N^{{K}}, \Phi
\label{VQ}
\ee
and impose the F-term conditions
\be
D_NW_{\rm mod}\,  =\,  0 \ \iff \  D_NW\, =\, K_N W_{\rm inf} \quad \quad \forall N^{{K}}
\ee
with $K_N = \p_NK$. Then, using the identities
\bea
\label{id1}
K^{\Phi\bar{\Phi}} K_{\bar{\Phi}} + \sum_{L} K^{\Phi\bar{N}^{L}} K_{\bar{N}^{L}} & = & 0 \\
\sum_{\a,\b = N^K, \Phi} K_\a K^{\a\bar{\b}} K_{\bar{\b}} & = & 4 
\label{id2}
\eea
and also imposing $\p_\Phi W_{\rm mod} =0$ we obtain that
\be
V_Q\, =\, e^K\left(K^{\Phi\bar{\Phi}} |\p_\Phi W_{\rm inf}|^2 + 4 |W_{\rm inf}|^2 \right) + {\cal O}(W_{\rm mod}^0)
\label{VQ2}
\ee

The second part of the potential is given by
\be
V_K\, =\, e^K \left(K^{T^a\bar{T}^b}D_{T^a} W D_{\bar{T}^b} \bar{W} \right) 
\label{VK}
\ee
Imposing the F-term condition
\be
D_{T^a} W_{\rm mod}\,  =\,  0 \ \iff \  D_{T^a} W\, =\, K_{T^a} W_{\rm inf}
\ee
which in particular implies $K_{T}  = 0$ and using the identity
\be
K_{{T}^a} K^{T^a \bar{T}} \, = \, 2i \,\im T
\label{id3}
\ee
we obtain
\bea \nonumber
V_K & = & e^K\left( K^{{T}\bar{T}} |\p_{T} W_{\rm inf} |^2 + 3|W_{\rm inf}|^2-  (2 \im T)^2 |\p_{T} W_{\rm inf}|^2 \right) \\
& & + {\cal O}(W_{\rm mod}^0)
\label{VK2}
\eea

Let us now add these two pieces of the potential and compare the full potential energy with that of the vacuum constructed from $W_{\rm mod}$. The result is
\bea
\nonumber
\Delta V & = & e^K \left(K^{\Phi\bar{\Phi}} |\p_\Phi W_{\rm inf}|^2 + (K^{{T}\bar{T}} + 4 (\re T)^2) |\p_{T} W_{\rm inf} |^2\right)\\
& + &   {\cal O}(W_{\rm mod}^0)
\label{diffpot}
\eea
By assumption, the above F-term conditions are compatible with $\Phi = \im T = 0$, so we can analyse (\ref{diffpot}) around that locus. Notice for instance that  $|\p_{T} W_{\rm inf} |^2 = |\Phi|^2$, and so at large values of the inflaton field $\re T$, $\Phi = W_{\rm inf} = 0$ is energetically favoured \footnote{In additon, taking $\Phi \neq 0$ will disturb the F-terms of the closed string fields that enter the moduli stabilisation superpotential $W_{\rm mod}$.}. 

At this locus the potential difference reduces to
\be
\Delta V\, =\, e^K K^{\Phi\bar{\Phi}} |T|^2
\label{Vinfsugra}
\ee
which is precisely the quadratic potential for $T$, and in particular for $b$, that we were aiming for from (\ref{Winf}). 

Let us now reconsider the caveats associated to F-term axion monodromy inflation in the scenario at hand. First, we have an inflaton candidate $b$ which only appears in the supergravity scalar potential through (\ref{Vinfsugra}). This quadratic term is special in the sense that it is suppressed by $K^{\Phi\bar{\Phi}}$, unlike the mass terms for the K\"ahler moduli that appear at $W_{\rm mod}$. One may then suppress the mass for the inflaton candidate with respect to other closed string moduli without making any further assumption for the superpotential and by simply decreasing the value for $K^{\Phi\bar{\Phi}}$ with respect to the closed string metrics like $K^{T^a\bar{T}^b}$. As hinted above, this may be done by placing the D6-brane creating the superpotential (\ref{Winf}) in a warped region of $\cM_6$ \footnote{See \cite{Franco:2014hsa} and references therein for warped throats in type IIA compactifications, as well as for a proposal to embed F-term axion monodromy inflation using them.}. Indeed, similarly to \cite{Marchesano:2008rg}, the effect of warping will be to enhance the constants $Q^K$ that appear in the K\"ahler potential (\ref{KQ}), and to which $K^{\Phi\bar{\Phi}}$ is inversely proportional. Hence by increasing the warping in a small region around the D6-brane one may decrease the inflaton mass in (\ref{Vinfsugra}), while at the same time keeping its kinetic term (that arises from integrals in the bulk) unaffected.

For instance, by direct dimensional reduction one can see that for the following choices of scales
\be
\cV_{\cM_6} \, \sim\, 10^3 \quad \quad \cV_{\Pi_3} \, \sim\, 10 \quad \quad g_s^2\, \sim \, 0.1
\label{scales}
\ee
with all volumes measured in the string frame and in units of the string scale, one gets a mass for the K\"ahler modulus $T$ in (\ref{Vinfsugra}) of the order \cite{toappear}
\be
m_{\rm inf} \sim Z_{\rm D6}^{-1/2} 10^{-3.5} M_{\rm pl}
\label{minf}
\ee
with $M_{\rm pl} \sim 10^{18}$ GeV the reduced Planck mass and $Z_{\rm D6}$ the warp factor around the D6-brane location. Hence, by choosing the value of $Z \sim 10^3$ one can get a realistic inflaton mass. 

As mentioned above, due to its shift symmetry (\ref{Vinfsugra}) is the only potential felt by the B-field axion $b =\re T$ defined in (\ref{binf}). Therefore it will acquire a mass of the order (\ref{minf}), being able to drive inflation along the trajectory $\{\re T = b_* \raw b_{\rm end}, \im T = 0\}$. However, before claiming that this the correct trajectory one must analyse the potential felt by the its saxion partner $\im T$. Since $\im T$ appears in the K\"ahler potential, it will also appear in the F-terms for the massive moduli stabilised by $W_{\rm mod}$ and therefore in the scalar potential generated before we added $W_{\rm inf}$. But this does not guarantee that it will acquire a high mass from the potential stabilising moduli. In fact, it was shown in \cite{Conlon:2006tq} that for supersymmetric vacua like the ones considered above $\im T$ will have a vanishing or tachyonic mass, becoming potentially unstable when an uplifting mechanism to de Sitter is invoked. Notice that the tachyonic mass is not a problem in our setting, since $\im T$  acquires a much larger and positive mass contribution (\ref{minf}), but it does show that the  mass term for $\im T$ is not bigger than that for the inflaton. Nevertheless, it may still happen that $\im T$ develops a quartic or higher term in the scalar potential which prevents it to acquire a large vev, and that it fixes it at vanishing value when we are rolling down with $\re T$. If that is the case, one may effectively have a system describing single field inflation, while if not one may have to perform a two field analysis like in \cite{Ibanez:2014swa,Bielleman:2015lka}.

Let us assume that we have such single field inflation system, leaving the two-field analysis for \cite{toappear}. Since the inflaton $b$ will take a trans-Planckian vacuum expectation value at the beginning of inflation, it is necessary to take into account Planck suppressed corrections to the quadratic potential described by supergravity. In our case this can be done by doing a dimensional reduction of the DBI action of the D6-brane, which sums over all $\a'$ corrections to the potential. One then finds that the quadratic potential is modified to \cite{toappear}
\be
V\, =\, \g \left( \sqrt{ 1 + \d \left(\frac{\phi_b}{M_{\rm pl}}\right)^2} -1 \right) M_{\rm pl}^4
\label{sqrtpotb}
\ee
where $\phi_b$ is the axion (\ref{binf}) canonically normalised and 
\bea
\label{cparam}
\g & \sim &  10^{-1}  g_s^3  \frac{\cV_{\Pi_3}}{ \cV_{\cM_6}^2}\\
\d^{-1} & \sim & 10^{-1} g_s^{-1} K_{\Phi\bar\Phi} K_{T\bar T} \cV_{\Pi_3} \cV_{\cM_6}
\label{aparam}
\eea
Again, one may get a realistic inflaton mass with a choice of scales like (\ref{scales}) and an appropriate choice of warp factor, obtaining that the above parameters range around $\sqrt{\d\g} \sim 10^{-5}-10^{-6}$ and $\d \sim 10^{-1}-10^{-3}$.

\begin{figure}[!h]
\begin{center}
\includegraphics[scale=0.22]{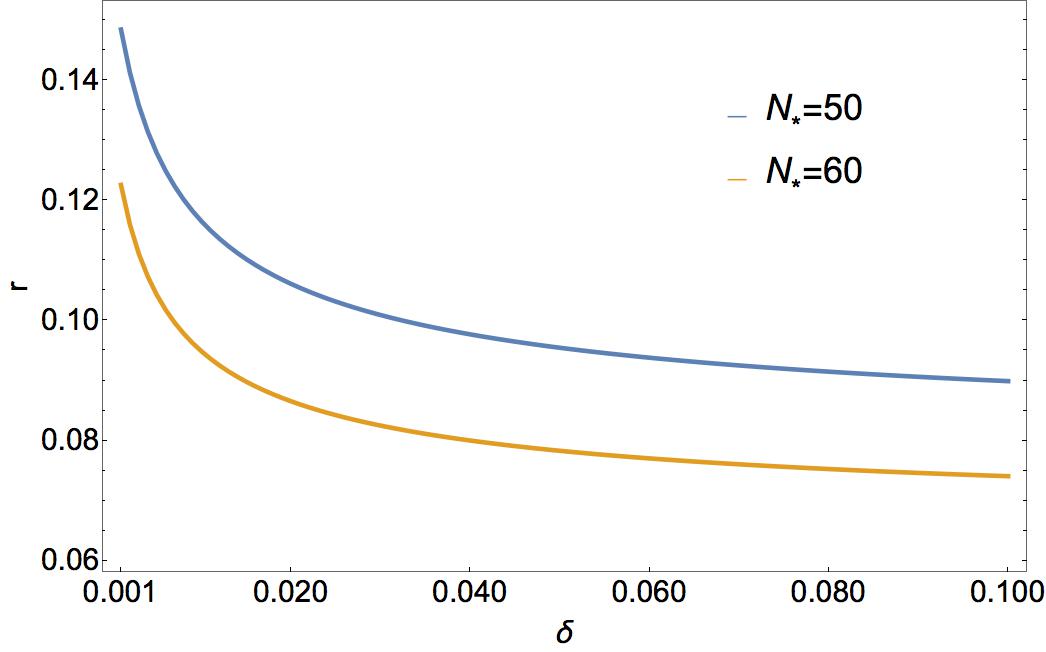}	
\end{center}	
\protect\caption{Tensor-to-scalar ratio obtained in terms of $\d$}
\label{r}
\end{figure}
\begin{figure}[!h]
\begin{center}
\includegraphics[scale=0.22]{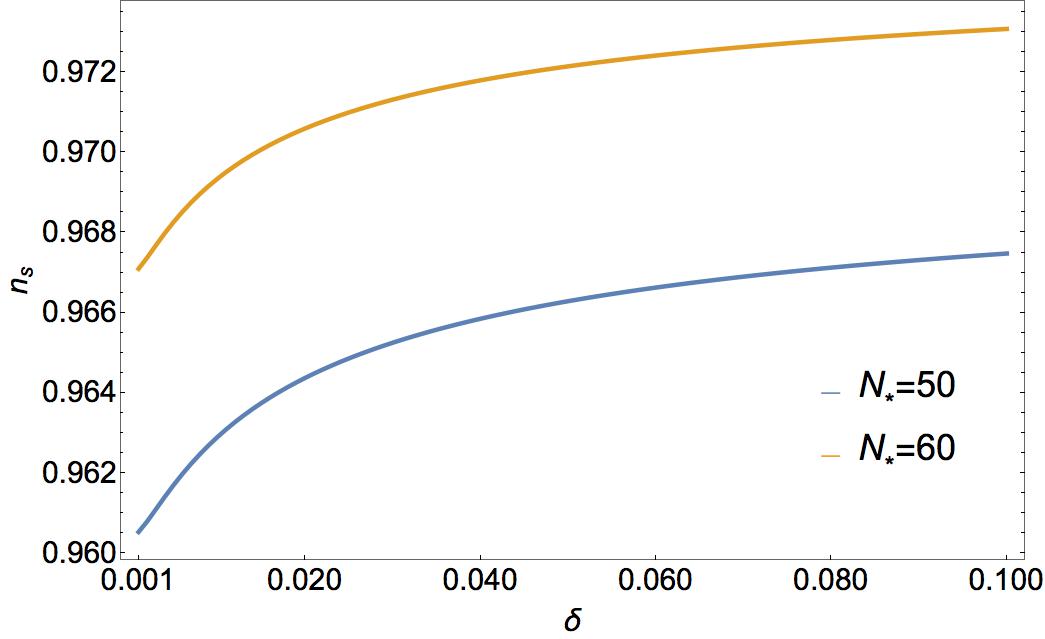}	
\end{center}	
\protect\caption{Spectral index in terms of $\d$}
\label{ns}
\end{figure}

Given this model of large single field inflation, one may compute the cosmological parameters associated to the range of parameters described above. One finds that slow-roll inflation typically occurs for $ 1.4 M_{\rm pl} < \phi_b  < 13-15 M_{\rm pl}$ for 60 efolds, and for $1.4 M_{\rm pl}< \phi_b < 12-14 M_{\rm pl}$ for 50 efolds, the upper limit $\phi_{b\,*}$ depending on the value of $\d$. In fact most cosmological parameters depend on $\d$, which interpolates between a model of quadratic chaotic inflation ($\d \sim 10^{-3}$) and a linear chaotic inflation ($\d \sim 10^{-1}$). 

\begin{figure}[!h]
\begin{center}
\includegraphics[scale=0.24]{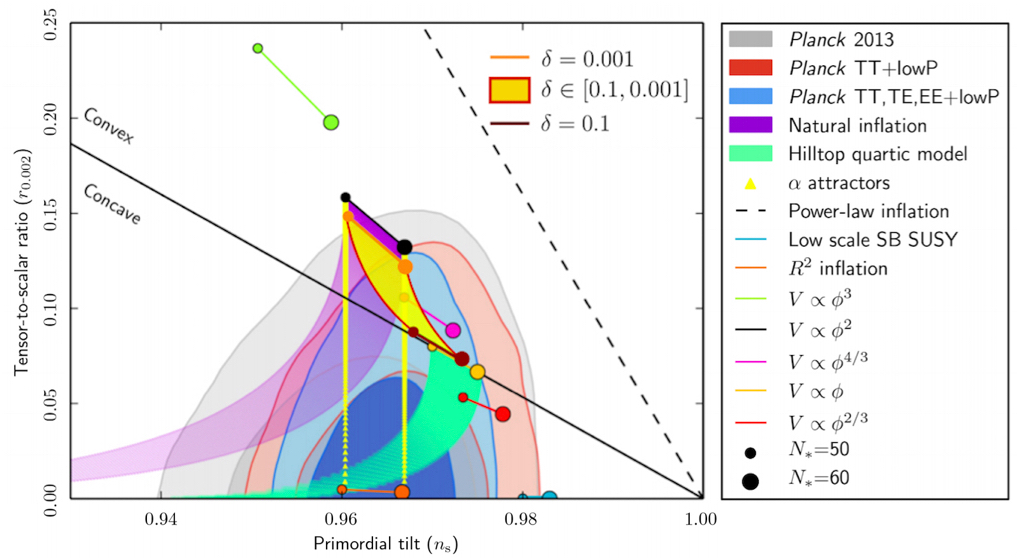}	
\end{center}	
\protect\caption{Primordial tilt $n_{s}$ vs $r$ given by Planck Collaboration. The yellow area shows the region covered by  $\d \sim 10^{-1} - 10^{-3}$.}
\label{nsrplanck}
\end{figure}

In figures \ref{r} and \ref{ns} we display the tensor-to-scalar ratio and the spectral index in terms of the parameter $\d$, for the number of efolds $N_* = 50$ and $N_* =60$. Finally, we plot one in terms of the other and compare with the plot given by Planck (2015) \cite{Planck} in figure \ref{nsrplanck}.

We leave a more detailed analysis of these quantities and higher order inflationary parameters for \cite{toappear}, but in general we find that the model satisfies all current constraints. It would be interesting  to build concrete examples of this type IIA scenario, as well as to explore similar string theory setups like type IIB with D7-branes. Finally, it would be interesting to see the effect of one-loop corrections to our moduli stabilisation scenario, to generalise it to consider some non-vanishing F-terms for massive moduli and to carry an analysis along the lines of \cite{Buchmuller:2014vda,Buchmuller:2015oma}. We hope to report on these issues in the future.

\vspace*{-.25cm}

\subsection{Acknowledgments}

We thank Luis Ib\'a\~nez for useful discussions.
This work has been supported by the grant FPA2012-32828 from the MINECO, the REA grant PCIG10-GA-2011-304023  (Marie Curie Action), the ERC Advanced Grant SPLE under contract ERC-2012-ADG-20120216-320421 and the Severo Ochoa grant SEV-2012-0249. D. E. and A.L. are supported through the FPI grants SVP-2014-068283 and SVP-2013-067949, respectively. F.M. is supported through the grant RYC-2009-05096. D.R is supported by a grant from the Max Planck Society.


\begin{thebibliography}{3}%
\makeatletter
\providecommand \@ifxundefined [1]{%
 \@ifx{#1\undefined}
}%
\providecommand \@ifnum [1]{%
 \ifnum #1\expandafter \@firstoftwo
 \else \expandafter \@secondoftwo
 \fi
}%
\providecommand \@ifx [1]{%
 \ifx #1\expandafter \@firstoftwo
 \else \expandafter \@secondoftwo
 \fi
}%
\providecommand \natexlab [1]{#1}%
\providecommand \enquote  [1]{``#1''}%
\providecommand \bibnamefont  [1]{#1}%
\providecommand \bibfnamefont [1]{#1}%
\providecommand \citenamefont [1]{#1}%
\providecommand \href@noop [0]{\@secondoftwo}%
\providecommand \href [0]{\begingroup \@sanitize@url \@href}%
\providecommand \@href[1]{\@@startlink{#1}\@@href}%
\providecommand \@@href[1]{\endgroup#1\@@endlink}%
\providecommand \@sanitize@url [0]{\catcode `\\12\catcode `\$12\catcode
  `\&12\catcode `\#12\catcode `\^12\catcode `\_12\catcode `\%12\relax}%
\providecommand \@@startlink[1]{}%
\providecommand \@@endlink[0]{}%
\providecommand \url  [0]{\begingroup\@sanitize@url \@url }%
\providecommand \@url [1]{\endgroup\@href {#1}{\urlprefix }}%
\providecommand \urlprefix  [0]{URL }%
\providecommand \Eprint [0]{\href }%
\providecommand \doibase [0]{http://dx.doi.org/}%
\providecommand \selectlanguage [0]{\@gobble}%
\providecommand \bibinfo  [0]{\@secondoftwo}%
\providecommand \bibfield  [0]{\@secondoftwo}%
\providecommand \translation [1]{[#1]}%
\providecommand \BibitemOpen [0]{}%
\providecommand \bibitemStop [0]{}%
\providecommand \bibitemNoStop [0]{.\EOS\space}%
\providecommand \EOS [0]{\spacefactor3000\relax}%
\providecommand \BibitemShut  [1]{\csname bibitem#1\endcsname}%
\let\auto@bib@innerbib\@empty
\bibitem [{Note1()}]{Note1}%
  \BibitemOpen
  \bibinfo {note} {See \cite {Font:2006na,Dudas:2014pva,Hayashi:2014aua} for
  other proposals to realise the superpotential (\ref {KYY}) in type II
  compactifications.}\BibitemShut {Stop}%
\bibitem [{Note2()}]{Note2}%
  \BibitemOpen
  \bibinfo {note} {In additon, taking $\Phi \not =0$ will disturb the F-terms
  of the closed string fields that enter the moduli stabilisation
  superpotential $W_{\protect \rm mod}$.}\BibitemShut {Stop}%
\bibitem [{Note3()}]{Note3}%
  \BibitemOpen
  \bibinfo {note} {See \cite {Franco:2014hsa} and references therein for warped
  throats in type IIA compactifications, as well as for a proposal to embed
  F-term axion monodromy inflation using them.}\BibitemShut {Stop}%
\end{thebibliography}%


\begin{thebibliography}{99}

\bibitem{Linde:1983gd} 
  A.~D.~Linde,
  Phys.\ Lett.\ B {\bf 129}, 177 (1983).

  \bibitem{Kawasaki:2000yn} 
  M.~Kawasaki, M.~Yamaguchi and T.~Yanagida,
  Phys.\ Rev.\ Lett.\  {\bf 85}, 3572 (2000)
  [hep-ph/0004243].
 
 \bibitem{Kallosh:2010xz} 
  R.~Kallosh, A.~Linde and T.~Rube,
  Phys.\ Rev.\ D {\bf 83}, 043507 (2011)
  [arXiv:1011.5945 [hep-th]].
   
\bibitem{Silverstein:2008sg} 
  E.~Silverstein and A.~Westphal,
  Phys.\ Rev.\ D {\bf 78}, 106003 (2008)
  [arXiv:0803.3085 [hep-th]].
  
  \bibitem{McAllister:2008hb} 
  L.~McAllister, E.~Silverstein and A.~Westphal,
  Phys.\ Rev.\ D {\bf 82}, 046003 (2010)
  [arXiv:0808.0706 [hep-th]].

\bibitem{Marchesano:2014mla} 
  F.~Marchesano, G.~Shiu and A.~M.~Uranga,
  JHEP {\bf 1409}, 184 (2014)
  [arXiv:1404.3040 [hep-th]].

\bibitem{Blumenhagen:2014gta} 
  R.~Blumenhagen and E.~Plauschinn,
  Phys.\ Lett.\ B {\bf 736}, 482 (2014)
  [arXiv:1404.3542 [hep-th]].
  
  \bibitem{Hebecker:2014eua} 
  A.~Hebecker, S.~C.~Kraus and L.~T.~Witkowski,
  Phys.\ Lett.\ B {\bf 737}, 16 (2014)
  [arXiv:1404.3711 [hep-th]].

\bibitem{Blumenhagen:2014nba} 
  R.~Blumenhagen, D.~Herschmann and E.~Plauschinn,
  JHEP {\bf 1501}, 007 (2015)
  [arXiv:1409.7075 [hep-th]].

\bibitem{Hayashi:2014aua} 
  H.~Hayashi, R.~Matsuda and T.~Watari,
  arXiv:1410.7522 [hep-th].

\bibitem{Hebecker:2014kva} 
  A.~Hebecker, P.~Mangat, F.~Rompineve and L.~T.~Witkowski,
  Nucl.\ Phys.\ B {\bf 894}, 456 (2015)
  [arXiv:1411.2032 [hep-th]].

\bibitem{Marchesano:2014iea} 
  F.~Marchesano, D.~Regalado and G.~Zoccarato,
  JHEP {\bf 1411}, 097 (2014)
  [arXiv:1410.0209 [hep-th]].

\bibitem{Font:2006na} 
  A.~Font, L.~E.~Ibanez and F.~Marchesano,
  JHEP {\bf 0609}, 080 (2006)
  [hep-th/0607219].

\bibitem{Dudas:2014pva} 
  E.~Dudas,
  JHEP {\bf 1412}, 014 (2014)
  [arXiv:1407.5688 [hep-th]].

\bibitem{thebook} 
  L.~E.~Ibanez and A.~M.~Uranga,
  Cambridge, UK: Univ. Pr. (2012) 673 p

\bibitem{Grimm:2004ua} 
  T.~W.~Grimm and J.~Louis,
  Nucl.\ Phys.\ B {\bf 718}, 153 (2005)
  [hep-th/0412277].
  
 \bibitem{Grimm:2011dx} 
  T.~W.~Grimm and D.~V.~Lopes,
  Nucl.\ Phys.\ B {\bf 855}, 639 (2012)
  [arXiv:1104.2328 [hep-th]].

\bibitem{Kerstan:2011dy} 
  M.~Kerstan and T.~Weigand,
  JHEP {\bf 1106}, 105 (2011)
  [arXiv:1104.2329 [hep-th]].

\bibitem{toappear} 
  D.~Escobar, A.~Landete, F.~Marchesano and D.~Regalado,
  arXiv:1511.08820 [hep-th].

\bibitem{Franco:2014hsa} 
  S.~Franco, D.~Galloni, A.~Retolaza and A.~Uranga,
  JHEP {\bf 1502}, 086 (2015)
  [arXiv:1405.7044 [hep-th]].


\bibitem{Marchesano:2008rg} 
  F.~Marchesano, P.~McGuirk and G.~Shiu,
  JHEP {\bf 0904}, 095 (2009)
  [arXiv:0812.2247 [hep-th]].

\bibitem{Conlon:2006tq} 
  J.~P.~Conlon,
  JHEP {\bf 0605}, 078 (2006)
  [hep-th/0602233].

\bibitem{Ibanez:2014swa} 
  L.~E.~Ibanez, F.~Marchesano and I.~Valenzuela,
  JHEP {\bf 1501}, 128 (2015)
  [arXiv:1411.5380 [hep-th]].
  
  \bibitem{Bielleman:2015lka} 
  S.~Bielleman, L.~E.~Ibanez, F.~G.~Pedro and I.~Valenzuela,
  arXiv:1505.00221 [hep-th].

\bibitem{Planck}
  P.~A.~R.~Ade {\it et al.} [BICEP2 and Planck Collaborations],
  Phys.\ Rev.\ Lett.\  {\bf 114}, 101301 (2015).

\bibitem{Buchmuller:2014vda} 
  W.~Buchmuller, C.~Wieck and M.~W.~Winkler,
  Phys.\ Lett.\ B {\bf 736}, 237 (2014)
  [arXiv:1404.2275 [hep-th]].

\bibitem{Buchmuller:2015oma} 
  W.~Buchmuller, E.~Dudas, L.~Heurtier, A.~Westphal, C.~Wieck and M.~W.~Winkler,
  JHEP {\bf 1504}, 058 (2015)
  [arXiv:1501.05812 [hep-th]].

\end{thebibliography}
\end{document}